%% file: ShovRescattering.tex
\documentclass[12pt,a4paper]{article}
\pdfoutput=1 

\usepackage[utf8]{inputenc}
\usepackage{amsmath}
\usepackage{graphicx}
\usepackage{amssymb}
\usepackage{mathrsfs}
\usepackage{mathcomp}
\usepackage{axodraw2}
\usepackage{enumerate}
\usepackage{cite}
\usepackage{xspace}
\usepackage{ifthen}
\usepackage{url}
\usepackage{booktabs}
\usepackage{multirow}
\usepackage{subcaption}
\usepackage{marginnote}
\usepackage{lineno}
\usepackage{axodraw2}

\include{common-defs}

\newlength{\abstwidth}
\begin{document} 
\sloppy

\pagestyle{empty}

\begin{flushright}
	LU-TP-22-50\\
	MCnet-22-12\\
\end{flushright}

\vspace{\fill}

\begin{center}
	{\Huge\bf Impact of string interactions on the space-time evolution of hadronic vertices}\\[4mm]
	{\Large Smita~Chakraborty and Leif~L\"onnblad} \\[3mm]
	{ \texttt{smita.chakraborty@thep.lu.se},  \texttt{leif.lonnblad@thep.lu.se}}\\[1mm]
	{\it Theoretical Particle Physics,}\\[1mm]
	{\it Department of Astronomy and Theoretical Physics,}\\[1mm]
	{\it Lund University,}\\[1mm]
	{\it S\"olvegatan 14A,}\\[1mm]
	{\it SE-223 62 Lund, Sweden}
\end{center}

\vspace{\fill}

\begin{center}
  \begin{minipage}{\abstwidth}
    {\bf Abstract}\\[2mm] 
		
    We investigate the space-time picture of string evolution and
    hadron production in a fully string-based model for high energy
    collisions involving heavy ions. We find that although the density
    strings is quite large at the time of hadronization in a central
    heavy ion collision, the initial overlap between them right after
    the collisions is not necessarily large. We also find that when
    including string--string interactions using the so-called
    \emph{shoving} model, the density of strings is decreased which
    should dampen the rapid increase in string tension in the rope
    hadronization with multiplicity that we found in a previous paper.
		
  \end{minipage}
\end{center}

\vspace{\fill}

\phantom{dummy}

\clearpage

\pagestyle{plain}
\setcounter{page}{1}

\tableofcontents

\section{Introduction}
\label{sec:intro}

The \angantyr \cite{Bierlich:2018xfw} model for modelling heavy ion
(HI) collisions in \pytppp implements a fairly simple procedure for
stacking nucleon--nucleon (\NN) sub-collisions on top of each other,
to build up full HI events. Each sub-collision is generated using the
full power of the \pytppp multi-parton interaction (MPI) framework
together with initial- and final-state parton showers. The
combined parton-level sub-events are then hadronized together with the
Lund string fragmentation model \cite{Andersson:1983ia}. Even though
there are no collective effects in this model it is able to adequately
describe multiplicities in both \pPb\ and \PbPb\ events at the LHC,
and even predict multiplicities in \XeXe \cite{ALICE:2018cpu}. This
begs the question, if it is possible that the colour degrees of
freedom generated in the initial stages on the perturbative level in
terms of colour connections (dipoles) between produced partons, can
survive the hot and dense environment of a HI collision in the form of
strings that then fragment into hadrons.

In a series of articles \cite{Bierlich:2014xba,Bierlich:2017vhg,
  Bierlich:2020naj,Bierlich:2022oja,Bierlich:2022ned} we have been
investigating possible effects of interactions between strings in a
dense environment, and have shown that such models may indeed give
rise to collective effects such as anisotropic flow and strangeness
enhancement, without the need of introducing a thermalised
quark--gluon plasma (QGP). In this article we take a step back and
investigate in more detail the space--time picture that arises from
these models.

Among the string interactions, string shoving \cite{Bierlich:2020naj}
and rope hadronization \cite{Bierlich:2022oja, Bierlich:2022ned} would
impact the final-state hadron yields the most in heavy-ion
collisions. The novelty of these mechanisms is based on the
equilibrium transverse extent of the colour-electric field ($E$) of
each string. Once each string is formed, the colour-electric field
spread transversely to reach an equilibrium width $R$, as established
in \citeref{Bierlich:2020naj}. The electric field then is approximate
by a Gaussian transverse shape,
\begin{equation}
\label{eq:col-electricfield}
E = N \text{exp} \left( - \frac{\rho^2}{2R^2}\right),
\end{equation}
where $\rho$ is the transverse distance from a center of a string
of radius $R$, in cylindrical coordinates. The effect of two strings' electric fields repelling
each other, would give rise to a net `shoving' effect, which would push either string pieces away. In our implementation,
this effect is modelled by looking at the
strings after the final radius has been reached, and then transferring
the push to primary hadrons by shifting their $p_\perp$ after
hadronization.

Also the hadronization process is affected by the interaction between
colour-electric fields of the strings. We have suggested that the
stings would form wider `colour-ropes' with larger
effective string tension \keff, than just the sum of the tension of
the individual strings. When such ropes hadronize, the higher \keff\
is released, which is available for tunnelling mechanism
\cite{Andersson:1983ia}, and therefore producing more strange
quarks. Thus, rope hadronization influences the strangeness yields in
collision systems.

In a recent work \cite{Bierlich:2022ned}, where we apply our rope
hadronization to \AA\ collisions, we find that rope hadronization by
itself enhances strangeness yields too much in central \PbPb\ events
as compared to data. We believe that this occurs due to an
overestimation of the string density at the time of hadronization in
case of heavy-ion collisions. As we noted in that publication, inclusion of string shoving
mechanism would produce a more accurate impact-parameter distribution
of strings at the time of hadronization. The $p_\perp$ pushes
generated due to shoving would dilute the system due to the dense
initial state generating larger shoving force.

In the current implementation, rope hadronization and string shoving
mechanisms are not completely compatible. In the perfect case, rope
formation would require the precise locations of the string pieces in
impact parameter. That would require pushing the string pieces with
the $p_\perp$ generated due to string shoving at each time step during
string evolution, but as mentioned above, in our implementation in
\citeref{Bierlich:2020naj}, the $p_\perp$ is only transferred to the
primary hadrons, formed from hadronization. Therefore, the rope
effects are somewhat approximate in all systems, and this mostly
affects the yields in \AA\ collisions, giving an overestimate of the
string density.

Further work on reproducing signals such as final-state collective
effects in \AA\ in \pythia would require hadronic rescattering
\cite{Sjostrand:2020gyg,Bierlich:2021poz} and string interactions to
work together. The effectiveness of the implementation will depend on
how accurately the string interactions are able to produce the primary
hadronic vertices as initial conditions to the hadronic
rescattering. In this publication, we investigate the impact of string
shoving and rope hadronization on primary hadron vertices.

The manuscript is organised as follows. In
\sectref{sec:spacetime-strng}, we describe the space-time evolution of
strings and their transverse overlaps in \pp\ and \AA\ collisions. In
\sectref{sec:corrctn-strng-int}, we present how the primary
hadron vertices are determined in \pythia, and discuss how these are
affected by the shoving and rope models.  Lastly, we present our
conclusions and further comments in \sectref{sec:conclusions}.

\section{Space-time evolution of strings}
\label{sec:spacetime-strng}

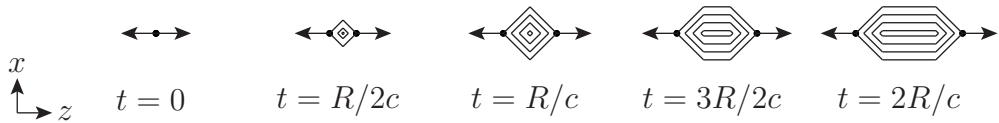
\begin{figure}
  \centering
  \input{figs/string.tex}
  
  \caption{Illustration on the time-evolution of the force field between
    a colour and an anti-colour charge produced in the same point and
    flying apart from each other along the $x$ axis with the speed of
    light. $R$ is the string radius.}
  \label{fig:string}
\end{figure}

The \angantyr model \cite{Bierlich:2018xfw} can be said to be a
straight forward extention of the multi-parton interaction model for
\pp in \pythia to HI collisions.
The model is based on an advanced Glauber
\cite{Glauber:1955qq,Miller:2007ri} calculation, which includes
so-called Glauber--Gribov \cite{Gribov:1968jf} corrections.  The
obtained nucleon--nucleon (\NN) sub-collisions, are produced with the
full \pytppp MPI machinery, and are basically stacked together and
hadronized. Some modifications are needed when one nucleon in one
nuclei interacts with several nucleons in the other. In this case only
one such sub-collision is treaded as \emph{primary} and is modelled as
a full \pp\ collision in \pythia, while the others are treated as
diffractive excitations similar to the wounded nuclei model
\cite{Bialas:1976ed} as described in detail in
\cite{Bierlich:2018xfw}.

As in \pythia, the final-state hadron multiplicity in \angantyr is
driven by multiple (semi--hard) scatterings among the partons of the
colliding nucleons. These are treated perturbatively, even if they can
be rather soft, which means that the scatterings are well
localised. In this picture, the partons are then connected by colour
lines, or \emph{dipoles}, and the colour field between the partons in
such a dipole is initially also well localised, but will spread out
with the speed of light ($c$) both longitudinally and transversely as
the partons fly apart. While the longitudinal extension will continue
to grow, the transverse extension will stop due to confinement, and we
get a string-like field with a constant string tension,
$\kappa\approx1$~GeV/fm, that will eventually break and form hadrons.

A simplified picture of the time-evolution of a single string piece
between a coloured and an anti-coloured parton flying apart with the
speed of light is given in \figref{fig:string}. Parameterising a point
along the string by the proper time, $\tau$, and hyperbolic angle, we
get in each time step, that the points where the radius reaches the
confinement
value, ($R$), will have $\tau c=R$.

\begin{figure}
  \centering
  \input{figs/initialgrowth}
  \caption{The evolution of the colour fields between partons in a
    sample PbPb event (top) and a sample pp event (bottom), both
    generated at $\sqrt{s_{{\tiny \NN}}}=2.76$~TeV. Each circle
    represents the position in impact parameter space of a colour
    dipole field that stretches across $z_{\text{lab}}=0$ at different
    times after the collision, $t_{\text{lab}}=0.1$~fm/c (left),
    $0.3$~fm/c (middle), and $0.5$~fm/c (right). The radius of each
    circles corresponds to the transverse extent of the colour field
    of the dipole as given by the proper time of the string field at
    that point.}
  \label{fig:initialgrowth}
\end{figure}
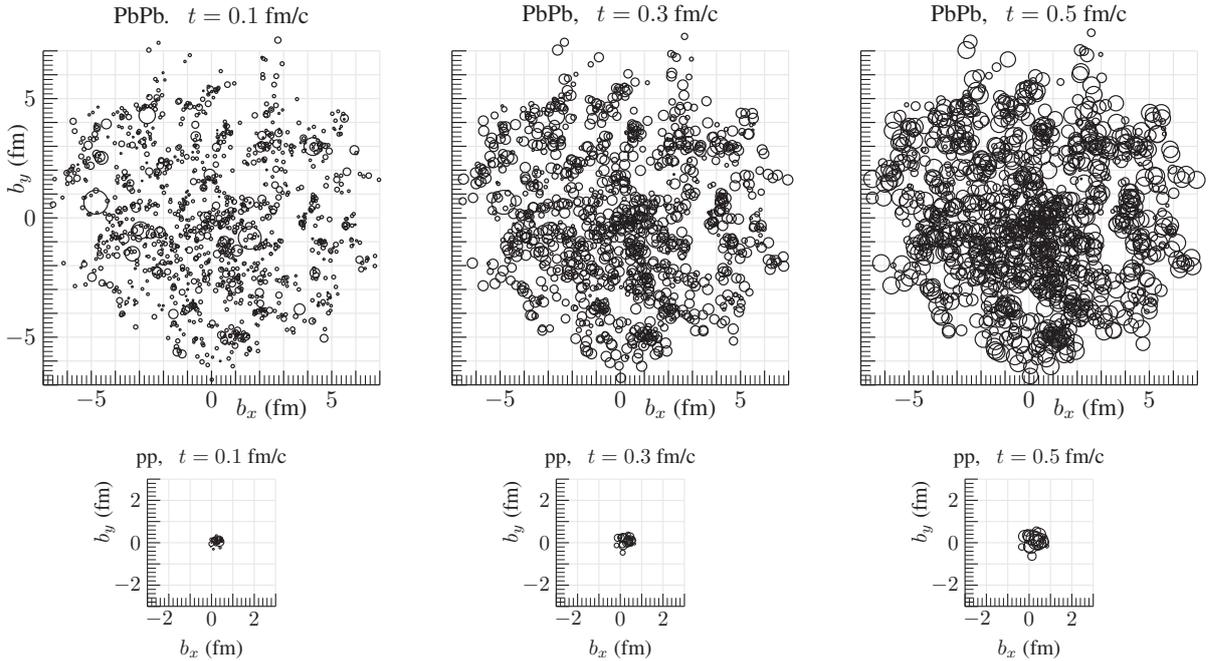

To illustrate the density of strings in a HI collision we have
generated a sample central \PbPb\ event at
$\sqrt{s_{{\tiny NN}}}=2.76$~TeV, shown in \figref{fig:initialgrowth}. The impact parameter is
$\approx0.2$~fm, resulting in a charged multiplicity in the central
pseudorapidity bin of $dN_{\text{ch}}/d\eta|_{\eta=0}\approx1750$. We then
look at the strings that span $z=0$ in the laboratory frame, and for
each such string we look at its size in different time steps (also in
the laboratory frame). For each of these we position in
\figref{fig:initialgrowth} a correspondingly sized circle in the impact
parameter plane.
To be more precise, the radius of each circle is given by the proper
time (multiplied by $c$)
at the
position along the string piece given $z=0$ and time in the laboratory
frame, limited from above by $R=0.5$~fm. In addition, to take into
account of how well localised the field were from the beginning, the
diameter of each circle is limited from below by
$\hbar c/p_{\perp\max}$, where $p_{\perp\max}$ is the largest
transverse momentum of the two partons spanning the field.

In the top left plot in \figref{fig:initialgrowth}, we see that initially, the
collision region is only sparsely
populated by the colour fields. Comparing to the lower left panel,
where we show a high multiplicity
($dN_{\text{ch}}/d\eta|_{\eta=0}\approx50$) \pp\ event at the same collision
energy ($\sqrt{s}=2.76$~TeV), there are very few regions in the \PbPb\
event where the colour fields are more densely packed.
At later times ($t=0.3$~fm/c in the middle panels), more of the
collision area is filled up by colour fields, and at $t=0.5$~fm/c
(right panels), almost the whole area is filled, and this is also when
the colour fields start to be confined to their maximum radius (0.5~fm
in this simulation). Most dipoles, however, have a tranverse momentum,
and due to the time-dilation, only some of them has reach their final
radius at $t=0.5$~fm/c.

Clearly the overlap between the colour fields in the \PbPb\ event
becomes quite large, which raises the question if the string picture
is really appropriate for heavy ion collisions. On the other hand, the
overlap is also quite high in the \pp\ collision, and we know that
\pythia is able to describe a vast range of hadronic final-state
observables in \pp. 
We want to see how far we can go with the string picture and, rather
than resorting to a hydro-dynamical approach with a quark--gluon
plasma,
we assume that the string degrees of freedom are still relevant for
hadronization.

Due to the large overlap among the strings, we need to worry about
possible string--string interactions, and in the following sections we
will discuss the space-time picture in our models for string shoving
and rope hadronization.

There is also a third effect that we will not discuss here, namely
colour reconnections. The assignment of colour connections between
partons in \pytppp\ is essential, not only for the string
fragmentation but also for the parton shower, which is based on the
dipole picture. The assignment of colours are, however, made on the
perturbative level in the $N_C\to\infty$ limit, and in a dense system,
there must be corrections to this. Indeed, already in the first
multi-parton interaction implementation in \pythia
\cite{Sjostrand:1987su} the concept of colour reconnections was
introduced where the colour connections between partons were allowed
to change before hadronization, in a way that favoured shorter
strings.

However, the colour reconnection models in \pytppp\ are based on a
pure momentum picture and does not take into account the space--time
separation between partons, so they are not suitable for heavy ion
collisions. And in the \angantyr model, there are therefore only
reconnections within each \NN\ collision, while reconnections between
them are not possible. We are currently working on a new reconnection
model that takes space--time constraints into account, but it is not
yet fully implemented.

\section{Corrections to hadronic vertices from string interactions}
\label{sec:corrctn-strng-int}

In this section, we present the 
shoving and rope models and how they affect the primary hadronic
vertices. In both of these models, the cumulated effect of interactions
between many stings are calculated by summing up pair-wise
interactions between string pieces. To calculate the interaction
between two string pieces, we use a special Lorentz frame, which we
call the parallel frame. Here, any two string pieces spanned between
two pairs of (massless) partons in any string system, will at any
given time be straight lines lying in parallel planes.\footnote{The
  full construction of the parallel frame is given in
  \citeref{Bierlich:2020naj,Bierlich:2022oja}.} This greatly
facilitates the calculation of the transverse shoving force, as well
as the increased string tension in the rope hadronization.

\subsection{String shoving}
\label{sec:shoving}

In our previous implementation of the shoving model
\cite{Bierlich:2020naj} we did not treat soft gluons properly. Gluons act like transverse
excitations or \emph{kinks} on a string. Since each gluon is connected
to two string pieces, it will lose energy to the strings twice as fast
as a quark. In any given reference frame, a gluon with energy $e$ will therefore have lost all its energy
after a time $t=e/2\kappa$. What happens then is that a new string
region is formed and will give a straight string piece starting from the point where
the gluon stops, expanding as if dragged out by the momenta of the
partons to which the stopped gluon is colour connected.
In general, a string spanned between a quark to an anti-quark via
series of gluons $(q_0, g_1, g_2, \ldots, g_{n-1}, \bar{q}_n)$ can
be treated as a series of string regions, or \emph{plaquettes}.
In these regions, we have the primary plaquettes spanned between the momenta $(p_i,p_{i+1})$
(where the momenta of the gluons is divided by two), corresponding to
the original dipoles, but also secondary ones spanned between
$(p_i,p_{i+2})$ with space--time vertices shifted by
$p_{i+1}/\kappa$. Similarly we get higher order plaquettes, spanned by
$(p_i,p_{i+n})$ with vertices offset by
$\sum_{j=i+1}^{i+n-1}p_j/\kappa$. This is explained in detail in
\citerefs{Sjostrand:1984ic} and \cite{Ferreres-Sole:2018vgo}.

In our updated shoving implementation, we now allow for all such
plaquettes. Just as for the primary dipoles, we can for each plaquette
look at any other plaquette in another string, go to the corresponding
parallel frame, and calculate the transverse force between them
there.

\begin{figure}
  \centering
  \vspace*{-5mm}
\includegraphics[width=0.5\textwidth]{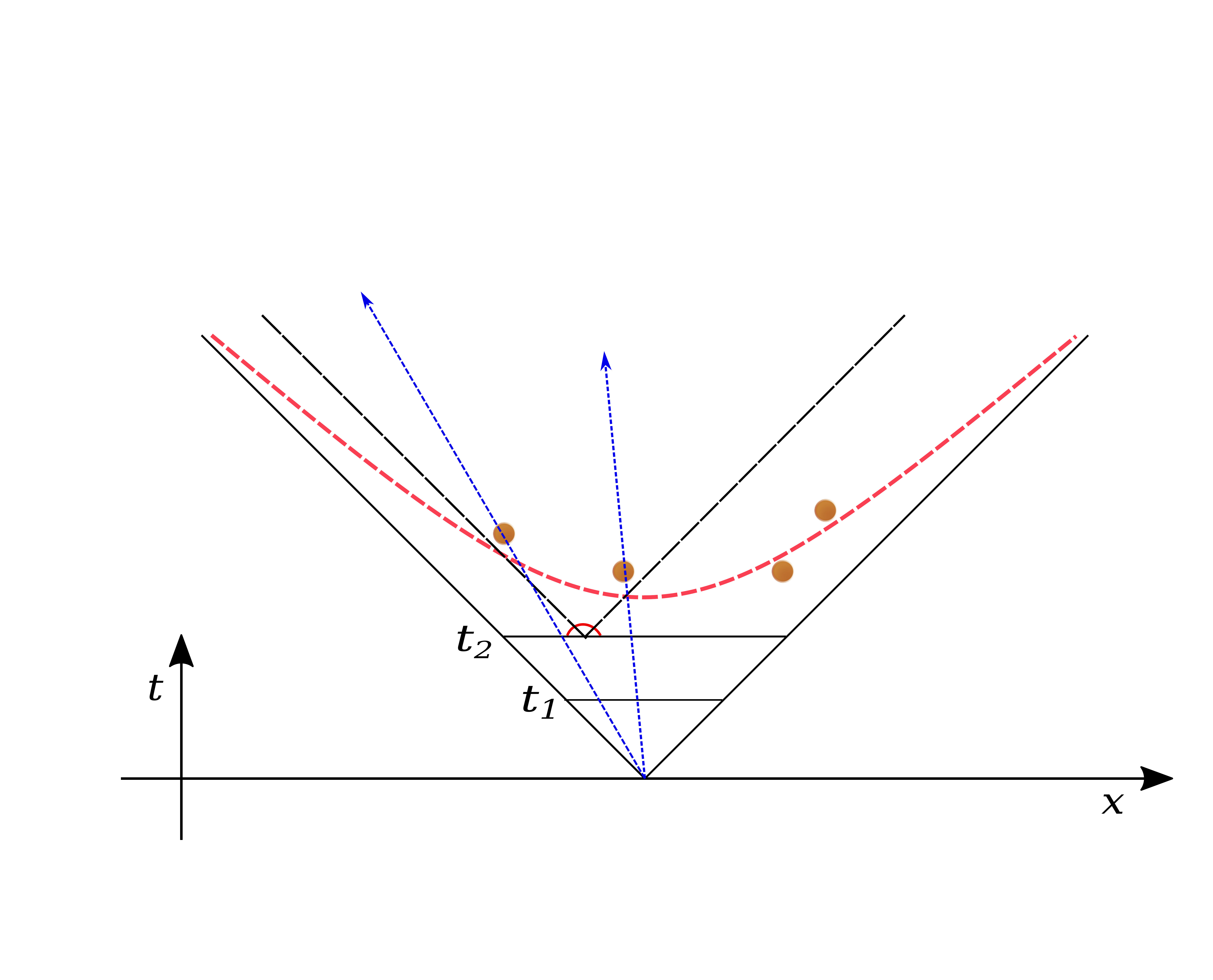}
\includegraphics[width=0.4\textwidth]{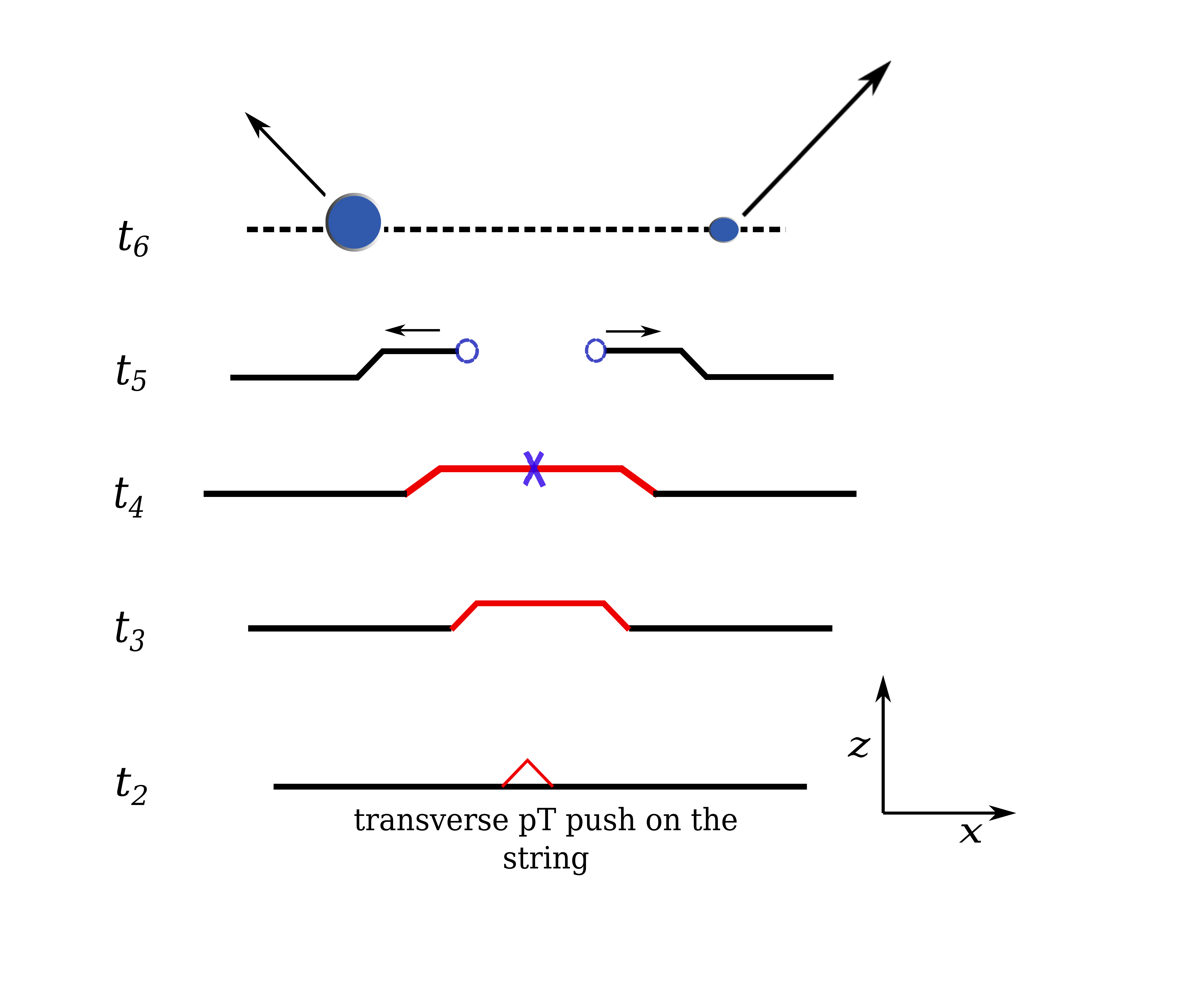}
\caption{Figures showing propagation of a localised push on a string
  moving along the $z$-axis. The left figure shows the space--time
  picture, and the right figure shows the deformation in of the string
  in the $z$ direction in different time steps.}
\label{fig:shoving-pT-dist}
\end{figure}

The shoving is implemented as discretised pushes with a small
($\sim 20$~MeV) transverse momentum, $\delta p_\perp$. In principle,
this would correspond to adding a small gluon to a string, but as its
energy would be very small it would stop almost immediately (after a
time, $\delta t=\delta p_\perp/2\kappa$). This would form a new plaquette which will
start to spread out with the speed of light in both directions of the
string, resulting in a shift of the string in the $z$
direction\footnote{In the parallel frame, the two stings pieces lie in
  planes parallel to the $x-y$ plane, moving in opposite directions
  along the $z$ axis (see \citeref{Bierlich:2020naj}).} with
$\delta_z=c\delta t$. This is illustrated in
\figref{fig:shoving-pT-dist} for one of the string pieces (the other
string will get a push in the opposite $z$ direction in a similar
way).

Implementing each new push with a new plaquette would give forbiddingly
complicated string configurations,\footnote{The complexity of our
  algorithm is already very high, requiring the construction of
  several millions parallel frames in a single central \AA\ event.}
and instead the transverse momentum of each push, which is localised
in the edges of the expanding region (with $\delta p_\perp/2$ on each
side), is transferred directly to the closest primary hadrons after
the hadronization.

Note that, when introducing secondary plaquettes, a push propagating
along the straight string piece may encounter a \emph{corner} between
two plaquettes and continue propagating in another
plaquette. Currently, such situations are only treated approximately,
assuming that the push will continue with in the same directions as in
the parallel frame where it is produced and the transverse momentum to
the primary hadron closest to that direction.

Previously, our implementation only considered the change in momentum
resulting from the shoving, but here we also want to study the
space--time picture. We note that the two hadrons receiving a
transverse-momentum push would be also pushed in space along the $z$
direction in the parallel frame. This is implemented by simply
shifting the production vertices assigned by the \pytppp\ string
fragmentation (see description in \sectref{sec:ropes} below) by
$\delta z$. In addition, any hadron produced between these two hadrons
along the string will be affected by the push and are also shifted by
$\delta z$. Note, however, that their momenta are not affected by the
push.

\begin{figure}
	\centering
	\includegraphics[width=0.30\textwidth]{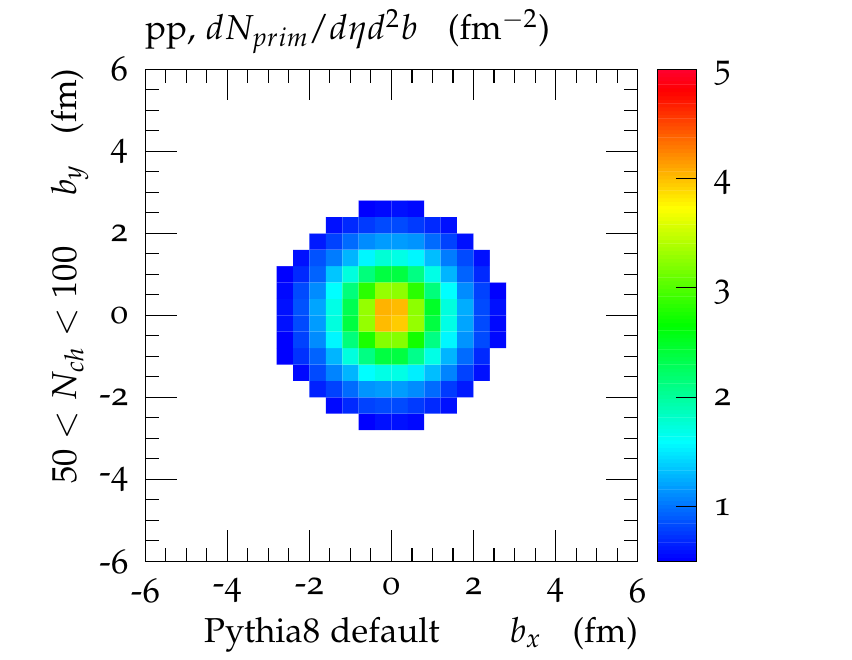}
	\includegraphics[width=0.30\textwidth]{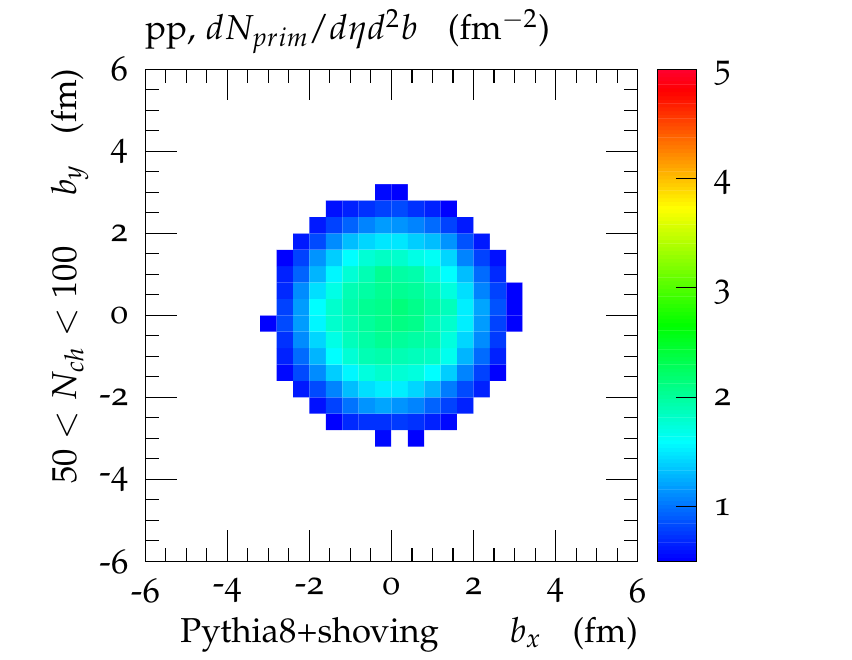}
	\includegraphics[width=0.32\textwidth]{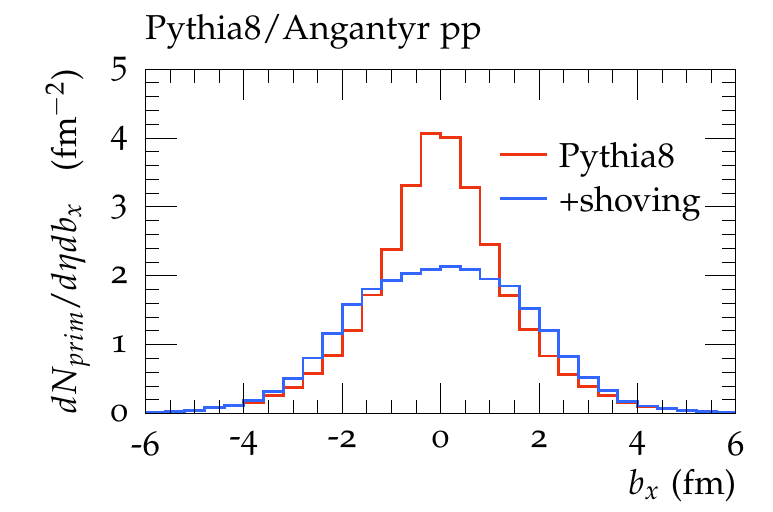}
	\includegraphics[width=0.30\textwidth]{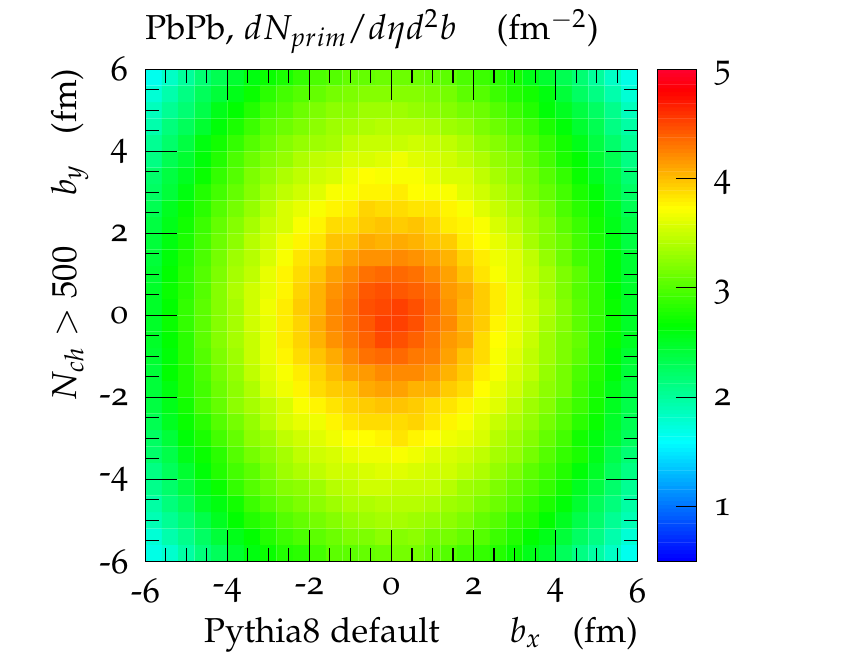}
	\includegraphics[width=0.30\textwidth]{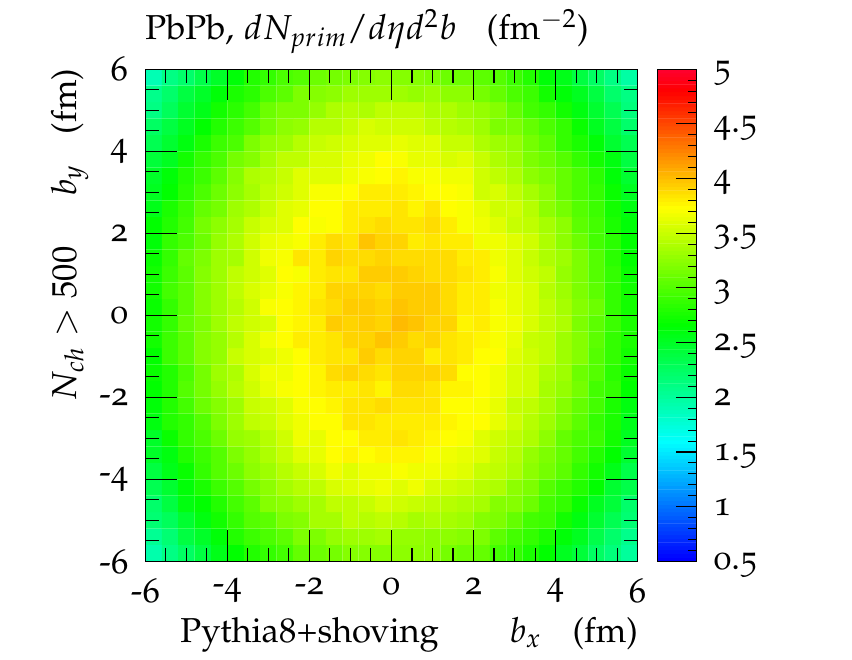}
	\includegraphics[width=0.32\textwidth]{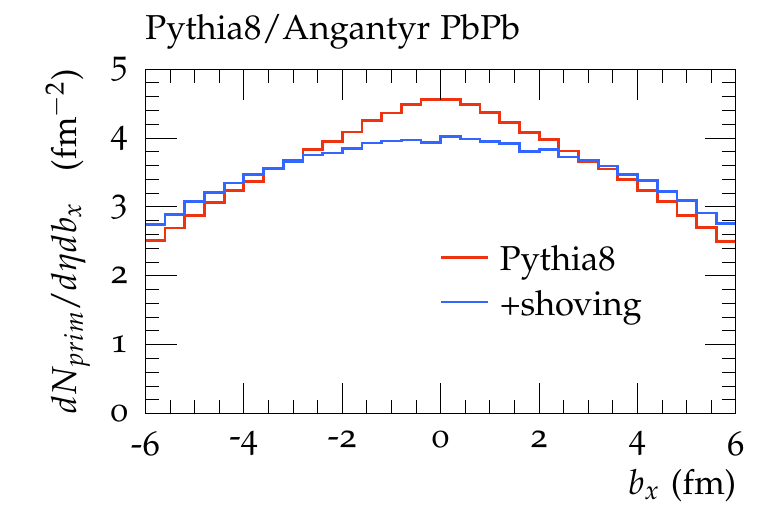}
	\caption{The distribution of vertices for primary hadrons in the
		shoving model for \pp\ collisions with central charged
		multiplicity, $50<dN_{\text{ch}}/d\eta|_{\eta=0}<100$ (top row), and for
		\PbPb\ collisions with $dN_{\text{ch}}/d\eta|_{\eta=0}>500$ (bottom
		row). The \pp\ events were generated at $\sqrt{s}=7$~TeV, and the
		\PbPb\ events at $\sqrt{s_{NN}}=2.76$~TeV. The left panels show a
		heat map, giving the number of primary hadrons with $|\eta|<0.5$
		per fm$^2$ in impact parameter space for \pytppp/\angantyr without
		shoving and the middle panels shows the same with shoving. The
		right panels compares the distributions with and without shoving in
		a slice with $|b_y|<0.5$~fm.}
	\label{fig:heatshove}
\end{figure}
String shoving therefore, affects the primary hadronic vertices
directly via their $z$ component in each parallel frame. This is what
we account for in this implementation and its effect is shown in
\figref{fig:heatshove}. Shoving would have the most effect in the most dense regions in a
collision. The correction due to shoving force would ``dilute'' the distribution
of primary hadrons transverse to the beam axis in the lab frame.

To show the effect of the shoving on the position of the primary
hadron vertices, we have generated a large sample of \pp\ and \PbPb\
events at 7~TeV and 2.76~TeV respectively, and looked at the positions
of the centrally produced ($|\eta|<0.5$) primary hadrons in impact
parameter space. In \figref{fig:heatshove}, we show in the top row the
resulting distributions with and without shoving\footnote{All basic
  \pytppp/\angantyr parameters have their default values, and the shoving model
  used a string radius of $R=0.5$~fm and a shoving strength of
  $g=0.25$.} for high-multiplicity \pp\ events (with charged
multiplicity, $N_{ch]}$ between 50 and 100 in the central rapidity
bin), and in the bottom row the same for high-multiplicity
($N_{\text{ch}}>500$) \PbPb\ events.

The first thing to note is that without shoving the high multiplicity
\pp\ events reaches almost as high densities of primary hadrons as the
very central \PbPb\ ones, which confirms what we already saw in
\figref{fig:initialgrowth}.\footnote{Note that the number of primary
  hadrons from a string is roughly one per unit rapidity.} With
shoving (middle column), we see that the vertices of primary hadrons
are more spread out compared to their distribution without shoving
(first column). The impact is most apparent for \pp\ collisions in the
first row, where the peak at $(0,0)$ is heavily dampened. This is
clearly seen in the right-most column, where for the $|b_y|<0.5$~fm
bins, the number density of primary hadrons are shown as a function of
$b_x$ for both with and without shoving. Also for \PbPb\ events the
hadronic vertices are more spread out with shoving included, but the
effect is not as dramatic. This is because the strings in the centre
are shoved from all sides, and therefor do not move as much.

\subsection{Rope hadronization}
\label{sec:ropes}

In \citeref{Bierlich:2022oja,Bierlich:2022ned} we presented a new rope
hadronization model for HI collisions in \angantyr, based on the
parallel frame technique described above. Depending on the transverse
separation between two string pieces at the time of hadronization, the
partons at the end of the strings combine to form higher
colour-multiplets. This would result in a higher effective string
tension, following lattice results and as established in our previous
works \cite{Bali:2000un, Bierlich:2022oja, Bierlich:2022ned}. When the
higher colour multiplet transitions to lower colour multiplets in a
string breaking, the energy from the higher string tension is
released. This results in an effective string tension, \keff, which is
higher than in a single string, increasing the possibility to produce
strange quarks in the tunnelling mechanism responsible for the
breaking. This would give rise to higher number of strange particles
as well as baryons in the final state. This \keff\ would, however,
also influence the production vertices of primary hadrons in various
stages, which we will describe below.

To calculate vertices of primary hadrons in \pythia, the relation
between the energy-momentum picture and space-time picture is used. In
the Lund model, the equation of motion of a string between a pair of
massless quark $q$ and antiquark $\bar{q}$, results in a linear
relation between space-time and energy-momentum:

\begin{equation}
\label{eq:lin-energymom-spacetime}
|\frac{dp_{x, q/\bar{q}}}{dt}| = |\frac{dp_{x, q/\bar{q}}}{dx}| = |\frac{dp_{q/\bar{q}}}{dt}| = |\frac{dp_{q/\bar{q}}}{dx}| =\kappa,
\end{equation}
where $\kappa$ is the string tension.  The location of a break-up
point on the string can be given by
$v_i=\frac{x_i^+ p^+ + x_i^- p^-}{\kappa}$, where $x_i^\pm$ are the
light cone fractions and $p^+$ ($p^-$) is the four-momentum of the $q$
($\bar{q}$). It is to be noted that these equations are not a function
of the width of the string from which the hadron is formed. In case of
a string with a radius $R$ the uncertainity of a hadronic vertex point
will arise in the transverse plane.

This effect is accounted for in the vertex calculation using a
Gaussian smearing\footnote{In \pytppp, a is called
  \texttt{HadronVertex:xySmear} controls the gaussian smearing and
  has a default value of 0.5~fm, which is the same value for $R$.}. We
will return to this effect while discussing the effect on \keff\ on
production vertices later.

Since a hadron is formed from two adjacent break-ups, the vertex
should be a function of each break-up point, say $v_i$ and
$v_{i+1}$. Since locating a hadronic vertex is not precise due to a
hadron's transverse extent, they are somewhat approximated. These
space-time locations of a hadronic vertex in \pytppp\ can therefore be
chosen in three different ways. The default definition is the ``middle
point'' in sampling the hadron vertex. This is given by:
\begin{equation}
\label{eq:intermediate-vertx}
v^h_i  = \frac{v_i + v_{i+1}}{2} \ \ \  \text{middle}.
\end{equation}
The ``early'' position is defined as the space-time point where the
backward light cones of the partons forming the hadrons cross. The
``late'' position is where the forward light cones cross:
\begin{align}
\label{eq:prim-vertices}
  v^h_{l,i} & = \frac{v_i + v_{i+1}}{2} + \frac{p_h}{2\kappa}\ \ \ \text{late},
              \nonumber\\
  v^h_{e,i} & = \frac{v_i + v_{i+1}}{2} - \frac{p_h}{2\kappa}\ \ \ \text{early}.  
\end{align}
For the detailed implementation, we refer the reader to
\citeref{Ferreres-Sole:2018vgo}.

For the purpose of this paper we have modified the vertex finding in
\pytppp, to take into account the increased string tension, \keff, in
the rope model. Since \keff\ varies along the string this is done
locally for each vertex.\footnote{It should be noted that \keff\ is
  only approximately localised along the string in our current
  implementation, and also the effects on the vertices are somewhat
  approximate.} It should be noted that \keff\ also affects the
transverse momentum of the of the $q\bar{q}$ pair in the tunnelling
mechanism, giving
$\sqrt{\langle k_\perp^2\rangle} \propto \sqrt{\keff}$, and this also
results in a corresponding scaling of the transverse momenta of the
hadrons. This means that the difference between \textit{early} and
\textit{late} for the transverse coordinates in
\eqref{eq:prim-vertices} would effectively be smaller than for the
longitudinal ones. In the following we will only use the default
\textit{middle} option in \eqref{eq:intermediate-vertx}.

\begin{figure}
  \centering
  \includegraphics[width=0.49\textwidth]{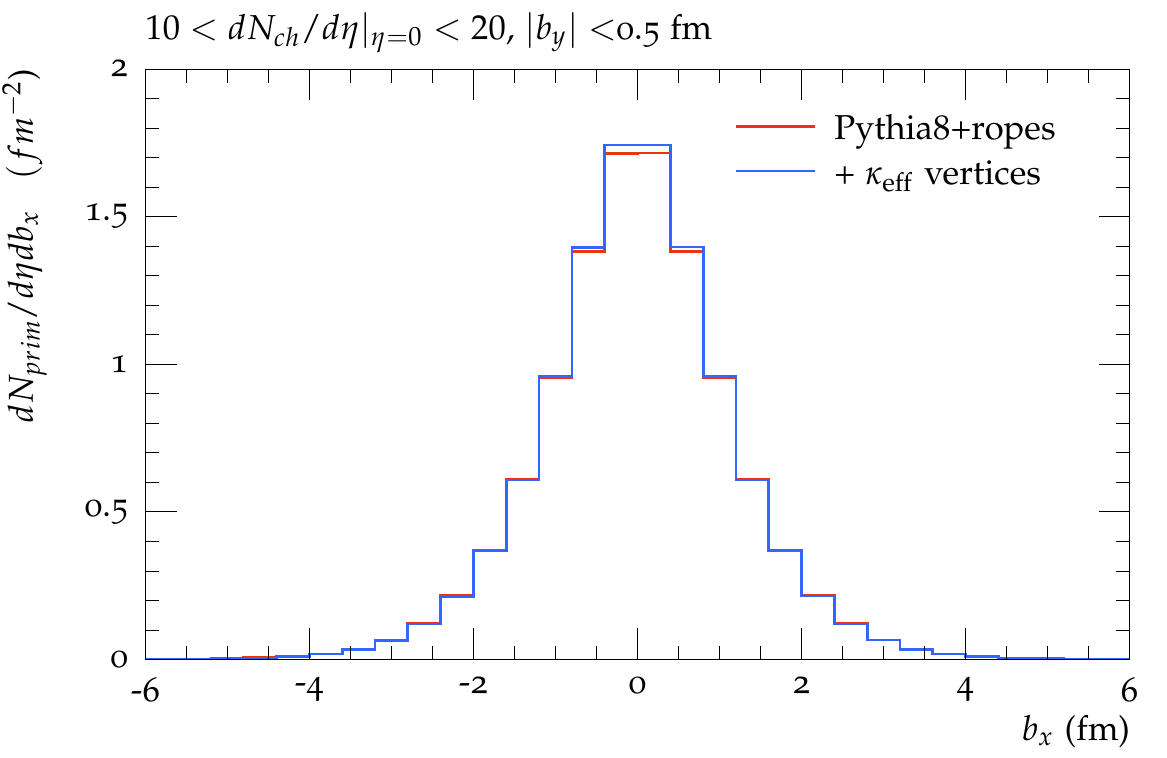}
  \includegraphics[width=0.49\textwidth]{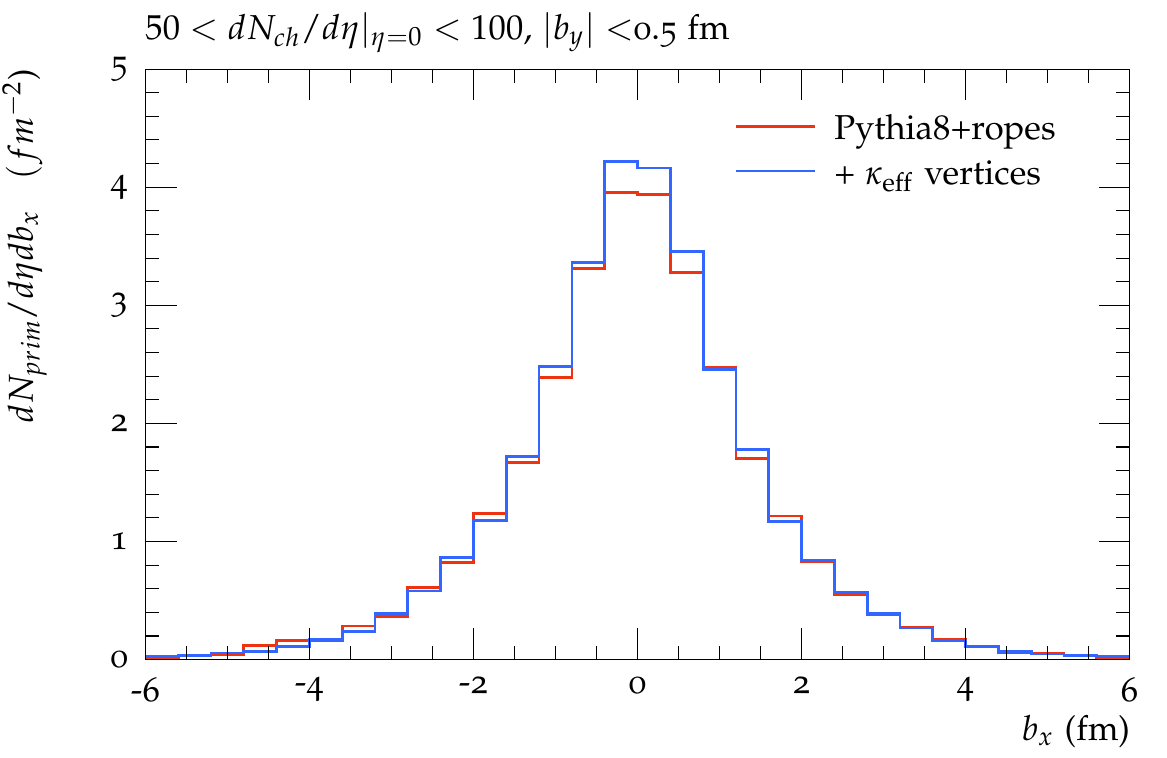}
  \caption{The distribution of vertices for primary hadrons in the
    rope hadronization model for \pp\ collisions with central charged
    multiplicity, $10<dN_{\text{ch}}/d\eta|_{\eta=0}<20$ (left), and
    $50<dN_{\text{ch}}/d\eta|_{\eta=0}<100$ (right). The events were
    generated at $\sqrt{s}=7$~TeV, and the distribution of vertex
    position along the impact parameter vector ($b_x$) is shown for a
    slice with $|b_y|<0.5$~fm. The blue line includes the effect of
    \keff\ on the vertex calculation in \pytppp, while the red line
    does not.}
  \label{fig:vkappa}
\end{figure}

When we include the modified \keff\ in calculation of the primary
hadron production vertices, the impact on the vertices proved to be
rather small. In \figref{fig:vkappa}, we show the effects in \pp\
collisions at $\sqrt{s}=7$~TeV, for a slice around $|b_y|<0.5$~fm in
impact parameter space, as a function of $b_x$. We show two
multiplicity bins $10 < \frac{dN_{\text{ch}}}{d\eta} < 20$ (left) and
$50 < \frac{dN_{\text{ch}}}{d\eta} < 100$ (right). As seen in the
figure, the effect is barely visible for the lower multiplicities due
to lower density of strings, but also for the higher multiplicities
the effect is small compared to the effects of shoving in
\figref{fig:heatshove}.

We have also studied the effect in HI collisions, and there it is even
smaller, since the vertex distribution in impact parameter is less
peaked than in the \pp\ case. Even for the highest multiplicities in
\PbPb\ collisions at 2.76~TeV, the effect (not shown here) is barely
visible, even though the densities of strings is larger than in \pp.

\section{Conclusion and outlook}
\label{sec:conclusions}

A correct description of primary hadron vertices in the Lund string
picture is essential to arrive at more consistent predictions from
string interactions in \pythia/\angantyr. This would help us determine
the contribution of such non-perturbative QCD effects on the
final-state observables both in small and large systems. As already
observed in \citerefs{Bierlich:2020naj, Bierlich:2022ned}, both
shoving and rope hadronization have non-trivial effects on the
final-state in dense systems. The effect is dependent on system size
and if accounted for correctly can reproduce QGP-like effects. This
would present different underlying mechanisms as origins of QGP-like
observations in both small and large systems.

We have shown here that in the \angantyr model, the initial occupancy
of colour fields from the MPIs is not large. As the fields grow
transversely they start shoving, giving rise to flow, and when they
finally hadronize the increased \keff\ gives rise to strangeness and
baryon enhancement. One should note that the rope hadronization and
string shoving models used in this publication is rather distinct to
CGC-Glasma picture (see \eg\ \citerefs{Lappi:2006fp,
  Venugopalan:2008xf, Gelis:2012ri}). While the Glasma initially
contains string-like features, it is assumed to be unstable and will
rapidly turn into a QGP. In our picture we instead assume that the
strings survive the dense environment and form ropes which then
fragments.

In this paper, we have shown that string shoving reduces the string
density, resulting in a smaller overlap between the stings forming
ropes. This reduces the effective string tension, which is used during
rope hadronization. That will in turn cause lower yield of strangeness
in dense environments. This would dampen the linear rise in the
strangeness yields for \AA\ using only rope effects as observed before
 \cite{Bierlich:2022ned}. Whether it will produce the
saturation behaviour as observed in data remains to be seen.

We have also observed that the increase of \keff\ in rope
hadronization, mainly affects the flavour of the hadrons and its
influence on the primary hadron vertices are small, especially when
compared to the effects of string shoving. While there are still some
caveats as discussed in the section \ref{sec:intro}, such as the
strings not being pushed in space--time before hadronization, we are
working towards a proper combination of both string shoving and
ropes. This would enable the string shoving to properly affect the
space-time overlaps between string, so that the dilution of strings
from the shoving can directly affect the \keff\ calculation in the
rope hadronization.

The primary hadron vertices are the main input to the hadronic
rescattering model in \pytppp, hence, one
could expect significant effects from the spreading of the vertices
due to shoving. Also, the rescattering has effects on the flavour
composition of the final state, and it is reasonable to assume that
there would be some interplay with rope model.

String shoving and hadronic rescattering together could provide an
enhanced final-state collectivity in both small and large
systems. Impact from hadronic rescattering in \angantyr ideally would be additive to that from string shoving, and they would build on
each other. This non-trivial effect needs proper correction to the
vertices of the primary hadrons, which is the work done in this
paper. By consistently combining shoving, ropes and rescattering, we
hope to achieve a consistent and complete picture for final-state
collectivity in all collision systems.

\section*{Acknowledgements}

We thank Gösta Gustafson and Christian Bierlich interesting
discussions and important input to this work.

This work was funded in part by the Knut and Alice Wallenberg
foundation, contract number 2017.0036, Swedish Research Council, contracts
number 2016-03291, 2016-05996 and 2017-0034, in part by the European
Research Council (ERC) under the European Union’s Horizon 2020
research and innovation programme, grant agreement No 668679, and in
part by the MCnetITN3 H2020 Marie Curie Initial Training Network,
contract 722104. 

\appendix

\bibliographystyle{utphys}
\bibliography{ShovRescatter}

\end{document}

%% file: common-defs.tex
\newcommand{\pythia}{P\protect\scalebox{0.8}{YTHIA}\xspace}

\newcommand{\pytppp}{P\protect\scalebox{0.8}{YTHIA}8\xspace}

\newcommand{\angantyr}{Angantyr\xspace}

\renewcommand{\eqref}[1]{eq.~(\ref{#1})\xspace}

\newcommand{\fig}[1]{\ref{#1}}
\newcommand{\figref}[1]{figure~\fig{#1}}

\newcommand{\citeref}[1]{ref.~\cite{#1}}
\newcommand{\citerefs}[1]{refs.~\cite{#1}}

\newcommand{\sect}[1]{\ref{#1}}
\newcommand{\sectref}[1]{section~\sect{#1}}

\def\text{\mathrm}
\def\eg{\emph{e.g.}\xspace}

\def\mrm#1{\mathrm{#1}}

\def\pp{\ensuremath{\mrm{pp}}\xspace}

\def\pPb{\ensuremath{\mrm{pPb}}}

\def\AA{\ensuremath{AA}\xspace}
\def\NN{\ensuremath{NN}}

\def\PbPb{\ensuremath{\mrm{PbPb}}}
\def\XeXe{\ensuremath{\mrm{XeXe}}}

\def\keff{\ensuremath{\kappa_{\mrm{eff}}}}

\def\ColText(#1,#2)[#3]#4{\Text(#1,#2)[#3]{#4}}

\def\showcommentsflag{0}
\newcommand{\showcomments}{\def\showcommentsflag{1}}

\newcounter{commentcounter}%
\newcommand{\comment}[1]{\ifnum\showcommentsflag > 0%
	\addtocounter{commentcounter}{1}%
	{{\Red{\ensuremath{\ddagger^{\arabic{commentcounter}}}}\Black{}}}%
	\marginpar{\raggedright\tiny\it{{\Red{\ensuremath{\ddagger^{\arabic{commentcounter}}}}} {#1}\Black{}}}
	\fi%
}
\newcommand{\commentdel}[2]{\ifnum\showcommentsflag > 0%
	\Red{\sout{#1}}\comment{#2}%
	\fi
}
\newcommand{\commentadd}[2]{\ifnum\showcommentsflag > 0%
	\comment{#2}\Red{#1}%
	\else
	#1
	\fi
}
\newcommand{\commentchange}[3]{\ifnum\showcommentsflag > 0%
	\Red{\sout{#2}}\comment{#3}\Red{#1}%
	\else
	#1
	\fi
}
\newcommand{\nocomment}[1]{\ifnum\showcommentsflag > 0%
	{\tiny\it\Red{\{#1}\}}
	\fi%
}
\newcommand{\nocommentdel}[1]{\ifnum\showcommentsflag > 0%
	\Red{\sout{#1}}%
	\fi
}
\newcommand{\nocommentadd}[1]{\ifnum\showcommentsflag > 0%
	\Red{#1}%
	\else
	#1
	\fi
}
\newcommand{\nocommentchange}[2]{\ifnum\showcommentsflag > 0%
	\Red{\sout{#2}}\Red{#1}%
	\else
	#1
	\fi
}

\showcomments

\makeatletter
\renewcommand\paragraph{\@startsection{paragraph}{4}{\z@}%
	{-3.25ex\@plus -1ex \@minus -.2ex}%
	{1.5ex \@plus .2ex}%
	{\normalfont\normalsize\bfseries}}
\makeatother

%% file: figs/string.tex
\newsavebox{\zerobox}
\savebox{\zerobox}(0,0)[cc]{
  \GOval(0,0)(1,1)(0){0}
  \GOval(0,0)(1,1)(0){0}
  \LongArrow(0,0)(-10,0)
  \LongArrow(0,0)(10,0)}

\newsavebox{\firstbox}
\savebox{\firstbox}(0,0)[cc]{
  \GOval(-5,0)(1,1)(0){0}
  \GOval(5,0)(1,1)(0){0}
  \LongArrow(-5,0)(-15,0)
  \LongArrow(5,0)(15,0)
  \Line(-5,0)(0,5)
  \Line(-5,0)(0,-5)
  \Line(5,0)(0,5)
  \Line(5,0)(0,-5)
  \Line(-2,0)(0,2)
  \Line(-2,0)(0,-2)
  \Line(2,0)(0,2)
  \Line(2,0)(0,-2)
  \Line(-0.5,0)(0,0.5)
  \Line(0.5,0)(0,0.5)
  \Line(-0.5,0)(0,-0.5)
  \Line(0.5,0)(0,-0.5)}

\newsavebox{\secondbox}
\savebox{\secondbox}(0,0)[c]{
  \GOval(-10,0)(1,1)(0){0}
  \GOval(10,0)(1,1)(0){0}
  \LongArrow(-10,0)(-20,0)
  \LongArrow(10,0)(20,0)
  \Line(-10,0)(0,10)
  \Line(-10,0)(0,-10)
  \Line(10,0)(0,10)
  \Line(10,0)(0,-10)
  \Line(-7,0)(0,7)
  \Line(-7,0)(0,-7)
  \Line(7,0)(0,7)
  \Line(7,0)(0,-7)
  \Line(-4,0)(0,4)
  \Line(4,0)(0,4)
  \Line(-4,0)(0,-4)
  \Line(4,0)(0,-4)
  \Line(-1,0)(0,1)
  \Line(1,0)(0,1)
  \Line(-1,0)(0,-1)
  \Line(1,0)(0,-1)}

\newsavebox{\thirdbox}
\savebox{\thirdbox}(0,0)[cc]{
  \GOval(-15,0)(1,1)(0){0}
  \GOval(15,0)(1,1)(0){0}
  \LongArrow(-15,0)(-25,0)
  \LongArrow(15,0)(25,0)
  \Line(-15,0)(-5,10)
  \Line(-15,0)(-5,-10)
  \Line(15,0)(5,10)
  \Line(15,0)(5,-10)
  \Line(-5,10)(5,10)
  \Line(-5,-10)(5,-10)
  \Line(-12,0)(-5,7)
  \Line(-12,0)(-5,-7)
  \Line(12,0)(5,7)
  \Line(12,0)(5,-7)
  \Line(-5,7)(5,7)
  \Line(-5,-7)(5,-7)
  \Line(-9,0)(-5,4)
  \Line(9,0)(5,4)
  \Line(-9,0)(-5,-4)
  \Line(9,0)(5,-4)
  \Line(-5,4)(5,4)
  \Line(-5,-4)(5,-4)
  \Line(-6,0)(-5,1)
  \Line(6,0)(5,1)
  \Line(-6,0)(-5,-1)
  \Line(6,0)(5,-1)
  \Line(-5,1)(5,1)
  \Line(-5,-1)(5,-1)}

\newsavebox{\fourthbox}
\savebox{\fourthbox}(100,60)[bl]{
  \GOval(-20,0)(1,1)(0){0}
  \GOval(20,0)(1,1)(0){0}
  \LongArrow(-20,0)(-30,0)
  \LongArrow(20,0)(30,0)
  \Line(-20,0)(-10,10)
  \Line(-20,0)(-10,-10)
  \Line(20,0)(10,10)
  \Line(20,0)(10,-10)
  \Line(-10,10)(10,10)
  \Line(-10,-10)(10,-10)
  \Line(-17,0)(-10,7)
  \Line(-17,0)(-10,-7)
  \Line(17,0)(10,7)
  \Line(17,0)(10,-7)
  \Line(-10,7)(10,7)
  \Line(-10,-7)(10,-7)
  \Line(-14,0)(-10,4)
  \Line(14,0)(10,4)
  \Line(-14,0)(-10,-4)
  \Line(14,0)(10,-4)
  \Line(-10,4)(10,4)
  \Line(-10,-4)(10,-4)
  \Line(-11,0)(-10,1)
  \Line(11,0)(10,1)
  \Line(-11,0)(-10,-1)
  \Line(11,0)(10,-1)
  \Line(-10,1)(10,1)
  \Line(-10,-1)(10,-1)}

\begin{picture}(360,40)(0,0)
  \LongArrow(0,0)(0,10)
  \LongArrow(0,0)(10,0)
  \Text(0,18)[c]{$x$}
  \Text(18,0)[c]{$z$}

  \put(50,30){\usebox{\zerobox}}
  \Text(50,5)[c]{$t=0$}
  \put(120,30){\usebox{\firstbox}}
  \Text(120,5)[c]{$t=R/2c$}
  \put(190,30){\usebox{\secondbox}}
  \Text(190,5)[c]{$t=R/c$}
  \put(260,30){\usebox{\thirdbox}}
  \Text(260,5)[c]{$t=3R/2c$}
  \put(330,30){\usebox{\fourthbox}}
  \Text(330,5)[c]{$t=2R/c$}

\end{picture}

%% file: figs/initialgrowth.tex
\begin{minipage}[c]{0.33\linewidth}
  \begin{picture}(170,180)(0,0)
    \SetScale{0.9}
    \SetOffset(80,80)
    \AxoGrid(-70,-70)(10,10)(14,14){LightGray}{0.5}
    \LinAxis(-70,-70)(70,-70)(14,5,5,0,0.5)
    \LinAxis(-70,70)(-70,-70)(14,5,5,0,0.5)
    \PText(-50,-76)(0)[]{$-5$}
    \PText(0,-76)(0)[]{$0$}
    \PText(50,-76)(0)[]{$5$}
    \PText(25,-80)(0)[]{$b_x$ (fm)}
    \PText(-73,-50)(0)[r]{$-5$}
    \PText(-73,0)(0)[r]{$0$}
    \PText(-73,50)(0)[r]{$5$}
    \PText(-80,10)(90)[l]{$b_y$ (fm)}
    \PText(0,85)(0)[]{PbPb.~~~$t=0.1$~fm/c}
    \input{figs/PbPb-1}
  \end{picture}
\end{minipage}%
\begin{minipage}[c]{0.33\linewidth}
  \begin{picture}(170,180)(0,0)
    \SetScale{0.9}
    \SetOffset(80,80)
    \AxoGrid(-70,-70)(10,10)(14,14){LightGray}{0.5}
    \LinAxis(-70,-70)(70,-70)(14,5,5,0,0.5)
    \LinAxis(-70,70)(-70,-70)(14,5,5,0,0.5)
    \PText(-50,-76)(0)[]{$-5$}
    \PText(0,-76)(0)[]{$0$}
    \PText(50,-76)(0)[]{$5$}
    \PText(25,-80)(0)[]{$b_x$ (fm)}
    \PText(0,85)(0)[]{PbPb,~~~$t=0.3$~fm/c}
    \input{figs/PbPb-3}
  \end{picture}
\end{minipage}%
\begin{minipage}[c]{0.33\linewidth}
  \begin{picture}(170,180)(0,0)
    \SetScale{0.9}
    \SetOffset(80,80)
    \AxoGrid(-70,-70)(10,10)(14,14){LightGray}{0.5}
    \LinAxis(-70,-70)(70,-70)(14,5,5,0,0.5)
    \LinAxis(-70,70)(-70,-70)(14,5,5,0,0.5)
    \PText(-50,-76)(0)[]{$-5$}
    \PText(0,-76)(0)[]{$0$}
    \PText(50,-76)(0)[]{$5$}
    \PText(25,-80)(0)[]{$b_x$ (fm)}
    \PText(0,85)(0)[]{PbPb,~~~$t=0.5$~fm/c}
    \input{figs/PbPb-5}
  \end{picture}
\end{minipage}\\[-10mm]
\begin{minipage}[c]{0.33\linewidth}
  \begin{picture}(170,120)
    \SetScale{0.8}
    \SetOffset(80,50)
    \AxoGrid(-30,-30)(10,10)(6,6){LightGray}{0.5}
    \LinAxis(-30,-30)(30,-30)(6,5,5,0,0.5)
    \LinAxis(-30,30)(-30,-30)(6,5,5,0,0.5)
    \PText(-18,-36)(0)[r]{$-2$}
    \PText(0,-36)(0)[]{$0$}
    \PText(20,-36)(0)[]{$2$}
    \PText(0,-50)(0)[]{$b_x$ (fm)}
    \PText(-33,-20)(0)[r]{$-2$}
    \PText(-33,0)(0)[r]{$0$}
    \PText(-33,20)(0)[r]{$2$}
    \PText(-50,00)(90)[l]{$b_y$ (fm)}
    \input{figs/pp-1}
    \PText(0,40)(0)[]{pp,~~~$t=0.1$~fm/c}
  \end{picture}
\end{minipage}%
\begin{minipage}[c]{0.33\linewidth}
  \begin{picture}(170,120)
    \SetScale{0.8}
    \SetOffset(80,50)
    \AxoGrid(-30,-30)(10,10)(6,6){LightGray}{0.5}
    \LinAxis(-30,-30)(30,-30)(6,5,5,0,0.5)
    \LinAxis(-30,30)(-30,-30)(6,5,5,0,0.5)
    \PText(-18,-36)(0)[r]{$-2$}
    \PText(0,-36)(0)[]{$0$}
    \PText(20,-36)(0)[]{$2$}
    \PText(0,-50)(0)[]{$b_x$ (fm)}
    \PText(-33,-20)(0)[r]{$-2$}
    \PText(-33,0)(0)[r]{$0$}
    \PText(-33,20)(0)[r]{$2$}
    \PText(-50,00)(90)[l]{$b_y$ (fm)}
    \input{figs/pp-3}
    \PText(0,40)(0)[]{pp,~~~$t=0.3$~fm/c}
  \end{picture}
\end{minipage}%
\begin{minipage}[c]{0.33\linewidth}
  \begin{picture}(170,120)
    \SetScale{0.8}
    \SetOffset(80,50)
    \AxoGrid(-30,-30)(10,10)(6,6){LightGray}{0.5}
    \LinAxis(-30,-30)(30,-30)(6,5,5,0,0.5)
    \LinAxis(-30,30)(-30,-30)(6,5,5,0,0.5)
    \PText(-18,-36)(0)[r]{$-2$}
    \PText(0,-36)(0)[]{$0$}
    \PText(20,-36)(0)[]{$2$}
    \PText(0,-50)(0)[]{$b_x$ (fm)}
    \PText(-33,-20)(0)[r]{$-2$}
    \PText(-33,0)(0)[r]{$0$}
    \PText(-33,20)(0)[r]{$2$}
    \PText(-50,00)(90)[l]{$b_y$ (fm)}
    \input{figs/pp-5}
    \PText(0,40)(0)[]{pp,~~~$t=0.5$~fm/c}
  \end{picture}
\end{minipage}